\documentstyle[12pt,epsf]{article}
\makeatletter
\def\@cite#1#2{{[{#1}]\if@tempswa\typeout
{IJCGA warning: optional citation argument
ignored: `#2'} \fi}}

\newcount\@tempcntc
\def\@citex[#1]#2{\if@filesw\immediate\write\@auxout{\string\citation{#2}}\fi
  \@tempcnta\z@\@tempcntb\m@ne\def\@citea{}\@cite{\@for\@citeb:=#2\do
    {\@ifundefined
       {b@\@citeb}{\@citeo\@tempcntb\m@ne\@citea\def\@citea{,}{\bf ?}\@warning
       {Citation `\@citeb' on page \thepage \space undefined}}%
    {\setbox\z@\hbox{\global\@tempcntc0\csname b@\@citeb\endcsname\relax}%
     \ifnum\@tempcntc=\z@ \@citeo\@tempcntb\m@ne
       \@citea\def\@citea{,}\hbox{\csname b@\@citeb\endcsname}%
     \else
      \advance\@tempcntb\@ne
      \ifnum\@tempcntb=\@tempcntc
      \else\advance\@tempcntb\m@ne\@citeo
      \@tempcnta\@tempcntc\@tempcntb\@tempcntc\fi\fi}}\@citeo}{#1}}
\def\@citeo{\ifnum\@tempcnta>\@tempcntb\else\@citea\def\@citea{,}%
  \ifnum\@tempcnta=\@tempcntb\the\@tempcnta\else
   {\advance\@tempcnta\@ne\ifnum\@tempcnta=\@tempcntb \else
\def\@citea{--}\fi
    \advance\@tempcnta\m@ne\the\@tempcnta\@citea\the\@tempcntb}\fi\fi}
\makeatother

\newcommand{\gsim}{\lower.7ex\hbox{$\;\stackrel{\textstyle>}{\sim}\;$}}
\newcommand{\lsim}{\lower.7ex\hbox{$\;\stackrel{\textstyle<}{\sim}\;$}}
\newcommand{\be}{\begin{equation}}
\newcommand{\ee}{\end{equation}}
\newcommand{\bea}{\begin{eqnarray}}
\newcommand{\eea}{\end{eqnarray}}

\def\baselinestretch{1}
\begin{document}
\catcode`@=11
\newtoks\@stequation
\def\subequations{\refstepcounter{equation}%
\edef\@savedequation{\the\c@equation}%
  \@stequation=\expandafter{\theequation}
  \edef\@savedtheequation{\the\@stequation}
  \edef\oldtheequation{\theequation}%
  \setcounter{equation}{0}%
  \def\theequation{\oldtheequation\alph{equation}}}
\def\endsubequations{\setcounter{equation}{\@savedequation}%
  \@stequation=\expandafter{\@savedtheequation}%
  \edef\theequation{\the\@stequation}\global\@ignoretrue

\noindent}
\catcode`@=12
\begin{titlepage}
\vspace{5mm}
\title{{\bf Large Scale Structure  from the Higgs fields of the
Supersymmetric Standard Model}}
\vskip2in
\author{  
{\bf M. Bastero-Gil$^{1}$\footnote{\baselineskip=16pt E-mail: {\tt
mbg20@pact.cpes.susx.ac.uk}}},
{\bf V. Di Clemente$^{2}$\footnote{\baselineskip=16pt E-mail: {\tt
vicente@hep.phys.soton.ac.uk}}}
and
{\bf S. F. King $^{2}$\footnote{\baselineskip=16pt E-mail: {\tt
sfk@hep.phys.soton.ac.uk}}}  
\hspace{3cm}\\   
 $^{1}$~{\small Centre of Theoretical Physics, University of Sussex,} \\ 
{\small Falmer, Brighton, BN1 9QJ, U.K.}
\hspace{0.3cm}\\
 $^{2}$~{\small Department of Physics and Astronomy, University of Southampton,} \\ 
{\small Highfield, Southampton, SO17 1BJ, U.K.}
\hspace{0.3cm}\\
} 
\date{} 
\maketitle 
\def\baselinestretch{1.15} 
\begin{abstract}
\noindent 
We propose an alternative implementation of the curvaton
mechanism for generating the curvature perturbations which does
not rely on a late decaying scalar decoupled from inflation 
dynamics. In our mechanism the supersymmetric
Higgs scalars are coupled to the inflaton in a hybrid inflation 
model, and this allows the conversion of the isocurvature perturbations  
of the Higgs fields to the observed curvature perturbations 
responsible for large scale structure to take place during reheating.
We discuss an explicit model which realises this mechanism in which
the $\mu$ term in the Higgs superpotential is generated after
inflation by the vacuum expectation value of a singlet field.
The main prediction of the model is that the spectral index
should deviate significantly from unity, $|n-1|\sim 0.1$.
We also expect relic isocurvature perturbations in neutralinos and
baryons, but no significant departures from gaussianity and no
observable effects of gravity waves in the CMB spectrum.

\end{abstract}

\thispagestyle{empty}
\vspace{2cm}
\leftline{\today}
\leftline{}

\vskip-20.3cm
\rightline{}
\rightline{SUSX-TH 02-024}
\rightline{SHEP 02-29}
\rightline{hep-ph/0211011}

\end{titlepage}


\setcounter{footnote}{0} \setcounter{page}{1}
\newpage
\baselineskip=20pt

\noindent

\section{Introduction}

According to the inflationary paradigm \cite{inflation}, 
the presently observed large
scale universe originated from a very small patch of space which underwent
a period of quasi-exponentially accelerated expansion known as inflation. 
In such an inflationary approach,
the very largest scales, which are now entering the
horizon, would have been in causal contact at very early times.
According to inflation this causal contact 
would have ceased some 50-60 e-folds before the end of inflation,
corresponding to so-called horizon exit to distinguish it from
the horizon entry as observed in the present epoch.
Inflation clearly provides an 
attractive explanation for why the cosmic microwave background (CMB)
radiation should have the same temperature from points in the sky
which would have been out of causal contact at the time of last 
scattering (the horizon problem). 
It also accounts for the observed flatness of the universe
$\Omega =1$ (the flatness problem), consistent with the 
CMB data \cite{cmb,boomerang,maxima,dasi,archeops,map}.

Another commonly stated success of inflation \cite{inflation}
is the fact that the observed primordial
density perturbations, which were first observed by COBE \cite{cobe} on  
cosmological scales just entering the horizon, and which are
supposed to be the seeds of large scale structure, could 
have originated from the quantum fluctuations of the inflaton field,
the scalar field which is supposed to be responsible for driving inflation.
In this scenario the quantum fluctuations of the inflaton field during the
period of inflation become classical perturbations at horizon exit,
giving a primordial curvature perturbation which remains
constant until the approach of horizon entry \cite{constancy}. The
advantage of this 
scenario is that the prediction for the nearly scale-invariant 
spectrum depends only
on the form of the inflaton potential, and is independent of
what goes on between the end of inflation and horizon entry.
The disadvantage is that it provides a strong restriction on models
of inflation. The price of such simplicity,
with one field being responsible for both inflation
and the primordial curvature 
perturbation often translates into a severe restriction on the 
parameters of the inflaton potential.
This often requires very small values for the couplings
and/or the masses which apparently renders many such theories unnatural.  

Recently it has been pointed out that in general it is unnecessary
for the inflaton field to be responsible for generating the
curvature perturbation \cite{lyth1,moroi}.
It is possible that the inflaton only generates a very small
curvature perturbation during the period of inflation,
which instead may result from the 
isocurvature perturbations of a curvaton field
which subsequently become converted into 
curvature perturbations in the period after inflation, but before
horizon entry \cite{silvia,linde,lyth1,moroi}. 
Isocurvature perturbations simply mean perturbations which
do not perturb the total curvature, usually because the curvaton
field contributes a very small energy density $\rho_{\sigma}$
during inflation. In the 
scenarios presented so far \cite{lyth1,moroi,others,moroi2,lyth2,lyth3}, 
the curvaton is assumed to be
completely decoupled from inflationary dynamics, and 
is assumed to be some late-decaying scalar which decays before the
time that neutrinos become decoupled. 

The reason why the curvaton is assumed to be late-decaying can be 
understood from the following argument \cite{lyth2}.
After reheating
the total curvature perturbation can be written as, 
\be
{\cal R} =  (1-f){\cal R}_{\rm r}+f{\cal R}_{\sigma}
\label{R}
\ee
where
\be
f=\frac{3\rho_{\sigma}}{4\rho_{\rm r}+3\rho_{\sigma}}
\label{f}
\ee
and (on unperturbed hypersurfaces on super-horizon sized scales)
\be
{\cal R}_i\approx -H\left(\frac{\delta \rho_i}{\dot{\rho}_i}\right)
\sim \left(\frac{\delta \rho_i}{{\rho}_i}\right)
\label{Ri}
\ee
where the curvaton density $\rho_{\sigma}$
and radiation density $\rho_{r}$, arising from the decay of the inflaton, 
each satisfy their own
energy conservation equations and each ${\cal R}_i$ remains constant
on super-horizon scales. The time evolution of ${\cal R}$ 
on these scales is then given by its time derivative,
\be
\dot{\cal R}\approx -H f(1-f)\frac{S_{\sigma {\rm r}}}{3}
\label{Rdot}
\ee
where $S_{\sigma {\rm r}}$ is called the entropy perturbation defined by
\be
S_{\sigma {\rm r}} \simeq -3 ({\cal R}_{\sigma}-{\cal R}_{\rm r})
\label{S}
\ee
The curvaton generates an isocurvature perturbation because
initially $\rho_{r}\gg \rho_{\sigma}$, and hence $f\ll 1$,
so that from Eq. (\ref{R}) the curvature perturbation
is dominated by ${\cal R}_{\rm r}$.
However as the universe expands and the scale factor $a$ increases 
while the Hubble constant $H$ decreases, the curvaton with mass $m$,
whose oscillations have effectively been frozen in by the large Hubble
constant, begins to oscillate and act as matter.
After this happens $\rho_{r}$ decreases as $a^{-4}$ while the
energy density in the curvaton field $\rho_{\sigma}$ has a slower 
fall-off as $a^{-3}$. Eventually the curvaton energy density $\rho_{\sigma}$
becomes comparable to the radiation density from the inflaton decay
$\rho_{r}$, and when this happens from Eq. (\ref{f}) we see that $f\sim 1$ 
and from Eqs. (\ref{Rdot}), (\ref{S}) this
leads to the growth of the total curvature perturbation
${\cal R}$ from the isocurvature perturbation 
${\cal R}_{\sigma}>{\cal R}_{\rm r}$.
This mechanism, which allows the curvaton isocurvature perturbations to become
converted into the total curvature perturbation, requires
the curvaton scalar to be late-decaying.

The main motivation behind the present paper is to propose an alternative
curvaton mechanism that removes the necessity for having a late-decaying 
scalar. To this end we make the following two observations:

(1) The first observation is that hybrid inflation \cite{hybrid,copeland}
involves more than one scalar field and ends by a phase transition in
which the fields involved move from the false, vacuum energy
dominated  potential minimum, towards the global minimum, and start
oscillating around it. In the process, the vacuum energy gets
redistributed among the fields such that their energy densities are
comparable. Thus any isocurvature perturbation in one of the 
hybrid inflation fields may be converted into curvature perturbations
during the on-set of reheating.
Note that the conversion does not take place until one of
the fields decays. Given that the inflaton field
is a singlet field, we can choose the curvaton field to be a flat
direction during inflation made of a pair of charged fields, where the
gauge interactions imply that the curvaton decays first.

(2) The second observation is that the
supersymmetric standard model provides a natural
candidate for such a pair of charged scalar fields: the two Higgs doublets.  
In order to allow a flat direction during inflation, we also require that
the $\mu$ term which usually couples the two Higgs doublets in the
minimal supersymmetric standard model must be set to zero.
However, providing this obstacle can be overcome,
it is natural to explore the attractive possibility that such  
Higgs fields are the seed of large-scale structure
in the Universe. The idea would be that the inflaton gives
a very small curvature perturbation, 
with the Higgs giving an isocurvature perturbation.
The coupling of the Higgs scalars to the inflaton then allows the
energy densities of the inflaton and the Higgs to become
of similar magnitude at the on-set of reheating,
allowing the conversion of isocurvature density perturbations to 
curvature perturbations.

In the remainder of paper we shall discuss a supersymmetric hybrid inflation
model based on these observations.
In this model the superpotential includes, in addition to the
two Higgs doublet superfields $H_u$ and $H_d$
of the supersymmetric standard model, also 
two gauge singlet superfields, 
one of them $\phi$ playing the role of the inflaton 
and the other $N$ coupling to the two Higgs doublets, 
effectively generating an effective $\mu$ term
from the coupling $NH_uH_d$ as
in the Next-to-Minimal Supersymmetric Standard Model (NMSSM) \cite{steve}. 
This happens after inflation when it gets a vacuum expectation value (vev) 
$N_0=<N>$. The $N$ superfield also couples to the inflaton $\phi$ as
$\phi N^2$, and therefore acts as a messenger allowing the Higgs fields 
to couple to the inflaton. 
In previous discussions of this model \cite{steve}
we assumed that during inflation the Higgs and $N$ fields
are held at the origin, but here we shall show that an alternative
inflationary trajectory is possible in which the inflaton
$\phi$ as well as the Higgs doublets are slowly rolling during
the inflationary epoch. The isocurvature perturbations of the Higgs fields
during inflation are converted into curvature perturbations
during the initial stages of the reheating process, when all these
fields begin to oscillate with comparable energy densities.

The layout of the rest of the paper is as follows.
The model is presented in section 2, where the evolution for
the background fields is studied. 
In section 3 we discuss the nature and evolution of the perturbations
during the epoch of inflation, and show that it is natural
for the inflaton to give the dominant curvature perturbation,
but still too small to account for the COBE value,
while the Higgs fields strongly contribute to the isocurvature
perturbations. In section 4 we describe the evolution of the
perturbations during reheating, after the Higgs has decayed 
and the Universe is made of a mixture of matter (the 
oscillating singlet fields) and radiation (Higgs decay products),
and show that the isocurvature perturbations are converted
into curvature perturbations. In section 5 
we comment on  some of the subtleties 
involved in the transition from inflation to reheating, and the issue
of preheating.  
The predictions of the approach are discussed in section 6, and in
section 7 we provide a summary.  

\section{The model: a new inflationary trajectory}

In this section we revisit the supersymmetric hybrid inflation
model based on the superpotential \cite{steve}:
\be
W= \lambda N H_u H_d - \kappa \phi N^2 \,, 
\label{superpot}
\ee
where $N$ and $\phi$ are singlet superfields, and $H_{u,d}$ are the Higgs
superfields, and $\lambda,\kappa$ are dimensionless couplings.
Other cubic terms in the superpotential are forbidding by imposing a
global $U(1)_{PQ}$ Peccei-Quinn symmetry. 
The superpotential in Eq. (\ref{superpot}) includes a linear
superpotential for the inflaton field, $\phi$, typical of hybrid inflation,
as well as the singlet $N$ coupling to Higgs doublets as in the NMSSM.
In the original version of this model \cite{steve} we assumed that
during inflation $N$ and $H_u$, $H_d$ were set to zero.
Here we discuss an alternative inflationary trajectory
in which these fields may take small values away from the origin,
consistent with slow roll inflation.

In order to satisfy the D-flatness
condition\footnote{Higgs fields are charged under $SU(2)_L\times
U(1)_Y$ interactions, 
such that there is what is called a D-term contribution in the
potential of the form $(g_2^2+g_1^2) ( H_u^2-H_d^2)^2/8$.
A ``D-flat'' direction is made of a combination of the
fields such that the D-term vanishes \cite{kolda}.}, 
we assume the values of the Higgs doublets during inflation to
be equal, $H_u=H_d=h$. Inflation takes place below the susy breaking
scale. Therefore, including the soft susy breaking masses, $m_\phi$,
$m_N$ and $m_h$, and trilinears $A_\kappa$, $A_\lambda$,  
the potential for the real part of the fields is: 
\bea
V&=&V(0)+ \frac{\kappa^2}{4}N^4 + \kappa^2(\phi-\phi_c^+)(\phi-\phi_c^-)N^2
+ \frac{1}{2} m_\phi^2 \phi^2   \nonumber \\
& & + \lambda N h^2 (\frac{A_\lambda}{\sqrt{2}}-\kappa \phi) +
\frac{\lambda^2}{4} h^4 + \frac{\lambda^2}{2} N^2 h^2 
+ \frac{1}{2} m_h^2 h^2  \,. 
\label{V}
\eea
In our previous work we set the Higgs terms to zero, $h=0$. 
Allowing the Higgs terms in Eq. (\ref{V})
we see that the term proportional to $\lambda N h^2$
can induce a non-zero value for $h$ and $N$ during inflation providing that 
the coefficient of this term
is negative $\frac{A_\lambda}{\sqrt{2}}-\kappa \phi<0$.
However, providing that the values of the fields $N$, $H_u$ and $H_d$ are
sufficiently small during inflation, as discussed later,
then inflation is controlled by the inflaton $\phi$,
as in the original model \cite{steve}, and will end when 
$\phi$ reaches one of its critical values $\phi_c^{\pm}$ given by:
\be
\kappa \phi_c^{\pm}= \frac{A_\kappa}{2 \sqrt{2}} \left( 1 \pm
\sqrt{1 - \frac{4 m_N^2}{A_\kappa^2}} \right) \,.
\label{phic}
\ee
These critical values correspond to the field dependent mass squared
$\kappa^2(\phi-\phi_c^+)(\phi-\phi_c^-)N^2$ changing sign and becoming
negative, signalling the end of inflation due to the destabilisation
of the inflationary trajectory, after which the fields 
$N$, $\phi$, $H_u$ and $H_d$ then approach their
global minimum and acquire their physical vevs $N_0$, $\phi_0$, $v_u$
and $v_d$.
For order of magnitude estimations, here on we will take $\phi_c \simeq
\phi_c^+ \simeq \phi_c^-$. 

The parameters of the potential in Eq. (\ref{V}) are selected by the 
following physical requirements. As discussed in \cite{steve}
the global PQ symmetry imposed on the superpotential Eq. (\ref{superpot})
solves the strong CP problem. When the PQ symmetry is 
broken by the vevs of the singlets, it leads to a very light axion
in the usual way \cite{axions}. The axion scale $f_a$ is set
by the vevs of the singlets, and it is constrained by astrophysical
and cosmological observations to be roughly in the window 
$10^{10} \,GeV \le f_a \le 10^{13} \,GeV$ \cite{axionbounds2,axionbounds3}.  
Given that $f_a \sim \phi_c \sim \phi_0\sim N_0 \sim 10^{13}\, GeV$
and the soft breaking term $A_k$ are expected to be of the order of
1 TeV, Eq. (\ref{phic}), this leads 
to a coupling constant of the order $\kappa \sim 10^{-10}$.
The same applies to $\lambda \simeq \kappa$, with $\mu=\lambda N_0 \sim 1 \,
TeV$. The smallness of $\lambda$ and $\kappa$ will be explained
in a companion paper \cite{vicente}.
Therefore, demanding a zero cosmological constant at the global
minimum requires the height  of the potential during inflation to be of
the order of 
\be
V(0)^{1/4} \simeq \frac{\sqrt{\kappa}}{2} \phi_c \simeq (10^8 \, GeV) \,,
\ee
with a Hubble parameter of the order of 
\be
H = \frac{V(0)^{1/2}}{\sqrt{3} m_P} \simeq 10\, MeV \,, 
\ee
where $m_P=M_{P}/\sqrt{8 \pi}=2.4\times10^{18} \,GeV$ 
is the reduced Planck mass.   

An important condition for inflation is that 
the inflaton mass $m_\phi$ (and also $m_h$) needs to
be small enough in order to ensure the slow-roll of the inflaton, as
determined by the slow roll
parameters defined in the standard way as \cite{inflation}: 
\begin{eqnarray} 
 \epsilon &\equiv& \frac{1}{2} m_P^2 
  \left( \frac{V'}{V} \right)^{2} \ll 1  \label{eq:slowepsilon} \\
 | \eta | &\equiv& \left| \frac{ m_P^2 V''}{V} \right| \ll 1
  \label{eq:sloweta}
\end{eqnarray} 
where $V' (V'')$ are the first (second) derivatives of the potential.
In this case we find the slow roll parameters for the $\phi$ field:
\bea
\eta_\phi &= & \frac{m^2_\phi}{3 H^2} < 1 \,,
\label{eta} \\
\epsilon_\phi &=& \frac{1}{2} \eta_\phi^2 \frac{\phi^2}{m_P^2} < 1 \,,
\label{epsilon}
\eea
where it is understood that these expressions are evaluated  
at some number of e-folds before the end of inflation. 
Eq. (\ref{epsilon})
is always fulfilled once $\eta_\phi <1$. This would require an inflaton
mass of the order of some MeVs at most, say $m^2_\phi \sim 0.1\,
H^2$. A similar constraint will also apply to the Higgs soft masses.
This is also compatible with the observational constraint
on the spectral index, $n=0.93\pm0.13$ \cite{cmb}, which in terms of
the slow-rolling parameters \cite{inflation}, 
$\mid n-1 \mid = 2
\eta_\phi - 6 \epsilon_\phi$, 
gives $\eta_\phi < 0.03$. The smallness of the inflaton and Higgs soft masses
which are orders of magnitude smaller than 
the typical soft breaking values assumed for the trilinear couplings,
is explained in a companion paper \cite{vicente}.

We have also included a constant vacuum energy contribution $V(0)$
in the potential in Eq. (\ref{V})
whose origin we do not specify. It can be explained
once the model is embedded in a supergravity (sugra) model \cite{steve2},
where the question of susy breaking can be addressed.
It should be pointed out however that sugra corrections 
generically give rise to scalar masses contributions of the order of
the Hubble parameter during inflation. 
This is the so-called $\eta$-problem \cite{eta}. Nevertheless, the
problem can be avoided \cite{copeland,steve2,heisenberg,stewart} 
by a suitable choice of the sugra model, i. e., the Kahler potential
and/or the superpotential for the scalar inflaton. Our approach is
that it is natural to have scalar masses  
of the order $H^2$ during inflation, but rather slightly smaller in
order to satisfy Eq. (\ref{eta}).  This would only require at most a
mild tuning of the parameters in the sugra Kahler potential. 

We now discuss an inflationary trajectory that will enable the
Higgs fields to be non-zero and slowly rolling during the inflationary epoch,
and hence to acquire an isocurvature perturbation.
The requirement that the $h,\phi$ fields be 
slow-rolling during inflation, means that
their effective field dependent masses must be smaller than $H^2$,
\bea
\frac{\partial^2 V}{\partial h^2} & \simeq & \lambda^2 (3 h^2 + N^2) + 2
\lambda N( \frac{A_\lambda}{\sqrt{2}}- \kappa \phi) + m_h^2 < H^2 \\
\frac{\partial^2 V}{\partial \phi^2} & \simeq & \kappa^2 N^2 + m_\phi^2 <
H^2 
\eea
This requirement implies that both the fields $N$ and $h$ must
take small values during inflation, with
\be 
\frac{N}{\phi_c} < 10^{-10} \,,\;\;\;\;\; \frac{h}{\phi_c} < 10^{-5}\,.
\label{small}
\ee
and also, as already remarked, must have soft masses of order an MeV
or less.

The $N$ field-dependent mass is much larger and positive, 
\be
\frac{\partial^2 V}{\partial N^2}  \simeq 
2\kappa^2(\phi-\phi_c^+)(\phi-\phi_c^-)\sim O({\rm TeV}^2) \,,
\ee
where we dropped the small term $3\kappa^2N^2$ using Eq. (\ref{small}).
Thus the $N$ field will oscillate  
with an amplitude damped by the expansion, following the evolution
equation:
\be
\ddot N + 3 H \dot N + \frac{\partial V}{\partial N}=0 \,,
\ee
where
\bea
\frac{\partial V}{\partial N} &\approx & (2 \kappa^2
(\phi-\phi_c^+)(\phi-\phi_C^-) + \lambda^2 h^2) N + \lambda h^2
(\frac{A_\lambda}{\sqrt{2}}- \kappa \phi ) \nonumber \\
&\approx & \omega_N^2(\phi) N +
\lambda h^2 (\frac{A_\lambda}{\sqrt{2}}- \kappa \phi ) \,,
\label{ddotn}
\eea   
and we have again dropped the small term $\kappa^2N^3$ in the first line
using Eq. (\ref{small}).
The frequency of the oscillations is then of the order $\omega_N(\phi)
\simeq \kappa \phi_c$.   

Apparently, as far as $N$ is concerned, we 
have violated the slow roll conditions. Since we have seen that $N$ does
not disturb the slow roll of $h$ and $\phi$ we should not be concerned about
this. Nevertheless, we shall see that in some sense $N$ can be
regarded as slowly  
rolling, according to the following argument.
The typical time scale during inflation is set by the rate of
expansion H, but the oscillations of $N$ are much faster than that. 
On the Hubble time scale as far as the motion of $N$ is concerned 
all that we can see is the average effect of the oscillations. The 
motion of $N$ is then given by the ``quasi'' constant term (``quasi'' in the
sense that is given by the other fields $\phi$ and $h$ that are
slow-rolling), plus the pure oscillatory term with an amplitude that
decays with the expansion. So when compared to the evolution of the
other fields, $\phi$ and $h$, in a few e-folds the oscillatory term
in $N$ will be averaged to zero, and if we require a local
minimum $\frac{\partial V}{\partial N}=0$ then we have effectively, 
from Eq. (\ref{ddotn}),
\be
N(t) \simeq - \frac{ \lambda h^2}{\omega_N^2(\phi)}
(\frac{A_\lambda}{\sqrt{2}}- \kappa \phi) \sim
\frac{h^2(t)}{\phi_c^2} \phi(t) \,,
\label{Nh}
\ee 
which relates the inflationary trajectory of the average value of $N$
to the slowly rolling $h,\phi$ fields. 
Therefore the oscillations of $N$ are not exactly around zero 
but its centre will move along with the inflaton and Higgs
field. Therefore in an effective sense
the $N(t)$ field will $also$ slowly roll along the
valley of minima controlled by the Higgs and the inflaton. 
Effectively the $three$ fields will follow a slow-roll trajectory, 
\bea
\dot \phi(t) \simeq - \eta_\phi H \phi(t) \,, \label{dotphi}\\
\dot h(t) \simeq - \eta_h H h(t) \,,\\
\dot N(t) \simeq - (\eta_\phi +2 \eta_h) H N(t) \label{dotN} \,, 
\label{velocity}
\eea
with $N(t) \ll h(t) \ll \phi(t)$, as in Eq. (\ref{small}) and
$\eta_h= m_h^2/(3 H^2)$, with $\eta_{\phi}= m_{\phi}^2/(3 H^2)$
as in Eq. (\ref{epsilon}).

\section{Evolution of the fluctuations during inflation}

During inflation with several light (slow-rolling) scalar fields
$\phi_\alpha$, the comoving curvature perturbation \cite{lukash} can
be written as \cite{gordon}    
\be
{\cal R} = 
H \sum_\alpha \left( \frac{ \dot \phi_\alpha}{\sum_\beta
\dot \phi_\beta^2} \right) Q_\alpha\,,  
\label{RQ} 
\ee
in terms of the gauge-invariant scalar field amplitude
perturbations, the Sasaki-Mukhanov variables $Q_i$ \cite{Qi},
\be
Q_\alpha = \delta \phi_\alpha + \frac{\dot \phi_\alpha}{H} \psi .
\ee
And for the entropy perturbations \cite{kodama} we have\footnote{This 
expression for $S_{\alpha \beta}$ in terms of the scalar field
perturbations holds as far as we can neglect the coupling between the
scalar, and the ratio $Q_\alpha/\dot \phi_\alpha \simeq Constant$,
both of which are a good approximation during inflation.}:
\be 
S_{\alpha \beta} \simeq  -3 H ( \frac{Q_\alpha}{\dot
\phi_\alpha} - \frac{ Q_\beta} {\dot \phi_\beta} ) \,, \label{SQ}   
\ee
These expressions may be compared to the 
(more intuitive) expressions given earlier in Eqs. (\ref{R}),
(\ref{S}), where${\cal R}_\alpha= H Q_\alpha/\dot \phi$.  More general
expression can be found in Appendix B, 
Eqs. (\ref{curvature}) and (\ref{entropy}). 
The initial curvature perturbation, Eq. (\ref{RQ}), will be
dominated by the field with the largest velocity, whilst the field
with the smallest velocity dominates the entropy perturbations in
Eq. (\ref{SQ}) \cite{silvia,starobinsky,sasaki,juan,gordon}. 

At first sight one might think that the $N$ field, with the smallest
velocity $\dot N$, according to Eq. (\ref{velocity}),
would give the largest isocurvature perturbation. 
However, since its evolution is  
controlled by the Higgs field, the ratio $Q_N/\dot N$ turns out to be
of the same order of magnitude as $Q_h/\dot h$, as discussed in Appendix A.
Thus the Higgs fields contribute strongly to the isocurvature perturbations,
and in addition dominates the evolution of the $N$ field during inflation
according to Eq. (\ref{Nh}).  

During inflation the curvature perturbation is dominated
by the field with the highest velocity, which according to 
Eq. (\ref{dotphi}) is the inflaton.
For a given quantity $A$, the spectrum is defined by \cite{inflation},
\be
P_A(k)\equiv \frac{k^3}{2 \pi^2} \langle |A|^2 \rangle \,,
\label{spectrum}
\ee
where $k$ is the comoving wavenumber. 
Using Eq. (\ref{RQ}) we find
\be
P_{{\cal R}}^{1/2} \simeq \frac{ 3 H_*^2}{m_\phi^2 \phi_*} \frac{ H_*}{2 \pi}
\simeq \frac{H_*}{2 \pi \eta_\phi \phi_*} \,,
\label{cobe}
\ee
where the subscript ``*'' denotes the time of horizon exit, $H_* a=k$. 
This is by far smaller than 
the required  COBE value $P_{{\cal R}}^{1/2}\simeq 5 \times 10^{-5}$
\cite{cobe}, unless we take 
$\kappa m_\phi \sim 10^{-18}\, GeV$, $i.e.$ an  
inflaton mass of the order of a few eV. Such a tiny value\footnote{  
The tiny mass could be due purely to  
radiative correction, $\delta m_\phi^2 \sim \kappa^2 (\kappa
\phi_c)^2$, if the soft-breaking mass is set to zero \cite{steve2}.}
is far smaller than the MeV value we need to satisfy the slow roll
conditions, as discussed previously. Therefore it is natural to 
suppose that the curvature perturbations are initially too small
to satisfy the COBE condition, but are generated by the conversion
of the isocurvature perturbations from the Higgs field, during
reheating, as discussed in the next section.

After horizon exit, the entropy and curvature perturbations evolve
independently until the end of inflation.
Following Ref. \cite{silvia}, by the time  
inflation ends the amplitude of the fluctuations can be given in terms
of their values at Hubble crossing ($a H_*=k$), 
\bea
{\cal R}|_i \sim \frac{H_*}{\dot \phi_*} Q_{\phi_*}  \simeq -\frac
{ Q_{\phi_*}}{ \eta_\phi \phi_*}\label{Rbc}\,,\\
S_{\phi h}|_i \sim \frac{8}{3} \frac{H_*}{\dot h_*} Q_{h_*} \simeq
-\frac{8}{3} \frac{Q_{h_*}} {\eta_h h_*} \label{sphihi}\,,
\eea
with 
\be
Q_{i*}= \frac{H_*}{\sqrt{2 k^3}}\,.
\ee
Clearly, given that $h \ll \phi$, we have ${\cal R} \ll S_{\phi
h}\mid_i$, so that the isocurvature perturbation from the Higgs
is nice and large, so that when it gets converted into the
curvature perturbation during reheating 
it can account for the COBE observation, as we now discuss.

\section{Evolution of the fluctuations during reheating}

The slow-rolling regime ends when the inflaton field reaches the
critical value $\phi_c$. At this point, the effective mass of the $N$
field changes sign, and it starts moving towards the global minimum of
the potential at $N_0 \sim O(\phi_c)$. The effective masses of $\phi$
and $h$ also start increasing, 
and they also move quickly towards their global
minimum values.  Around the global minimum, the effective masses $\bar{m}_i^2$
for all the fields are of the same order of magnitude,
$\bar{m}_{\phi} \simeq \bar{m}_{N} \simeq \bar{m}_{h}\sim \kappa^2
\phi_c^2 \sim O(1\, TeV)$. Therefore, they all start  
oscillating with similar frequencies and  $amplitudes$ of the order of
$\phi_c$. In this way, the vacuum energy dominating during inflation is
equally redistributed among the three fields, $\rho_\phi \sim \rho_N
\sim \rho_h$, 
with the energy density of the oscillating fields behaving as matter. 

This will be the situation until the fields decay and
transfer their energy density to radiation. Due to the smallness of
the singlet couplings, they are 
very long-lived, with a decay rate of the order $\Gamma_{N,\phi} \simeq
\kappa^2 \bar{m}_{N,\phi} \sim O(10^{-17}\, GeV)$, where 
$\kappa \sim \lambda\sim 10^{-10}$ whose smallness is explained in 
\cite{vicente}. 
Reheating lasts until the singlets completely decay, roughly around
the time $H \simeq \Gamma_\phi$.
On the other hand, the
Higgs field decays much faster through its gauge interactions,
with $\Gamma_h \simeq \alpha_W \bar{m}_h \sim O(10 \, GeV)$. 
Because of this, the decay of the Higgses can be considered as practically
instantaneous when compared to that of the singlets. 
However, given that $\Gamma_h/\bar{m}_h \sim \alpha_W \sim
0.01$, they will still have time to oscillate several times before decaying.
Notice also that $\Gamma_h \gg H$ at the beginning of the oscillating
phase. From that point of view, the decay of the Higgs can also be
considered ``instantaneous'': the Higgs decay before the energy
densities in the oscillating fields have been redshifted, so that through
the decay a fraction of the vacuum energy dominating during inflation
is transferred into radiation. Given that during
the decay we can neglect the effect of the expansion, we will also
assume that the fluctuations in the Higgs field at the end of
inflation are converted into those of the radiation fluid, with
$S_{\phi r}= S_{\phi h}|_i$ as the initial condition during reheating.

Therefore, during the reheating era, we are left with a two component
fluid, made of a mixture of matter $\rho_\phi$ ($\phi$ 
and $N$ oscillating) and radiation $\rho_r$ (Higgs decay
products), but such that initially $\rho_\phi(0) \simeq \rho_r(0)$.
The evolution equations for the system can be written as 
\bea
\dot \rho_{\phi} + 3 H \rho_{\phi} + \Gamma_\phi \rho_{\phi}=0 
\label{rhophidot} \,,\\  
\dot \rho_{r} + 4 H \rho_{r} - \Gamma_\phi \rho_{\phi}=0 
\label{rhordot}\,.   
\eea
It will take a while for the initial large fraction of radiation to
feel the effect of the inflaton decay products, and at the beginning
it is redshifted as usual like 
$\rho_{r} \propto a^{-4}$, with $\rho_\phi \propto a^{-3}$ and $a$
being the scale factor. It is only later when the contribution of the
decay products become comparable to the already present radiation that
the system is fully coupled, and we have $\rho_r \propto
a^{-3/2}$. The decay is completed by the time $H \simeq \Gamma_\phi$,     
and we are left again with radiation which redshifts in the usual way
$\rho_r \propto a^{-4}$. 

For comoving scales $k$ well outside the Hubble radius, $k \ll a H$,
the evolution equations for the fluctuations during reheating can be
approximated by\footnote{The exact equations  for a two-component fluid 
obtained from Ref. \cite{kodama} are given in Appendix B for
completeness. They also include the equations for the relative
velocity perturbations $V_{\alpha \beta}$. In Appendix C we give the
equations and approximate 
solutions for the scenario we are considering. In this case, in order
to understand the qualitative behaviour of the numerical solution is
enough to consider the evolution equations for ${\cal R}$ and
$S_{\alpha \beta}$, without including $V_{\alpha \beta}$,
Eqs. (\ref{rphir2}) and (\ref{sphir2}) in Appendix C.} by 
\bea
\frac{d {\cal R}}{d \ln a} &\simeq&  \frac{ 8 \rho_\phi \rho_r}{(3
\rho_\phi+4\rho_r)^2}S_{\phi r}  
\label{drlna} \,,\\
\frac{d S_{\phi r}}{d \ln a} &\simeq& 
-\frac{ 12\rho_r}{3\rho_\phi+4\rho_r} S_{\phi r} 
- \frac{\Gamma_\phi}{H} \frac{\rho_\phi}{\rho_r}\left(\frac{3
\rho_\phi+\rho_r}{4 \rho_r + 3 \rho_\phi} \right) S_{\phi r} ,.
\label{dslna}
\eea
Thus, at the beginning of reheating when $\rho_r \simeq
\rho_\phi$, the large isocurvature perturbations coming from the Higgs fields
in Eq. (\ref{sphihi}) which dominates the entropy perturbations,
acts as a strong source for the curvature
perturbations ${\cal R}$. Thus the curvature perturbations ${\cal R}$
quickly grow to become of the same order of magnitude as the 
initial entropy perturbation, and afterwards remains
practically constant. 

The transfer from entropy perturbations to curvature perturbations
is illustrated in Fig. (\ref{plot1}), based on a numerical integration of the
exact evolution equations for the perturbations given in Appendix C,
Eqs. (\ref{rphir}-\ref{vphir}), together with
Eqs. (\ref{rhophidot}-\ref{rhordot}), with the boundary conditions in   
Eqs. (\ref{Rbc}) and (\ref{sphihi}). 
From the numerical integration, we find asymptotically,
\be
{\cal R} \simeq \frac{1}{5} S_{\phi h}\mid_i  \label{Rf} \,.
\ee
Hence, using Eqs. (\ref{sphihi}), (\ref{qi}), we end with the
spectrum of adiabatic perturbations 
\be
P^{1/2}_{\cal R} \sim \frac{1}{5} P^{1/2}_{S_{\phi r} \mid_i} \sim
\frac{8}{15 \eta_h h_*} P^{1/2}_{Q_{h_*}} \sim  \frac{8}{15 \eta_h}
\frac{H_*}{2 \pi h_*}\,,
\ee
which for $\eta_h \sim 0.03$, $H_* \simeq 10^{-2} \,GeV$ and $2 \pi h_*
\simeq 1 \, TeV$ gives the right order of magnitude as measured by COBE. 

\begin{figure}[t]
\epsfysize=10cm \epsfxsize=10cm \hfil \epsfbox{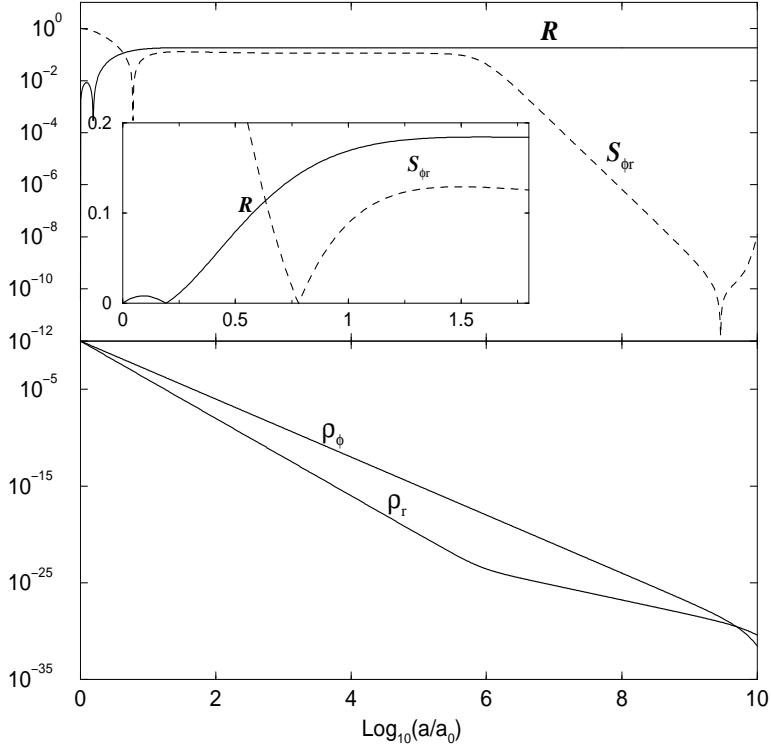} \hfil
\caption{\footnotesize In the upper panel we show the
evolution of the curvature ${\cal R}$ (solid line) and
entropy $S_{\phi r}$ (dashed line) perturbations during
reheating. Both values are normalised to the initial value of $S_{\phi r}$. 
We have taken $\Gamma_\phi=10^{-17}
GeV^{-1}$, $V(0)^{1/4}=10^{8} \, GeV$, 
$\phi/h=10^{10}$ and $\eta_\phi=\eta_h$ . 
In the inset in the upper panel we show a blow-up of the initial region
on a linear vertical scale.
In the lower panel we show the evolution of
the energy densities $\rho_\phi$ and $\rho_r$, normalised to their
initial values.}
\label{plot1}
\end{figure}

The entropy perturbation also remains practically constant until the
terms due to the decay of the singlets $N$ and $\phi$
become relevant, and the radiation
starts behaving as $\rho_r \propto a^{-3/2}$; the same as the
ratio $\rho_\phi/H$. From Eq. (\ref{dslna}), the evolution for $S_{\phi
r}$ is now approximately given by:  
\be
\frac{1}{S_{\phi r}} \frac{d S_{\phi r}}{d \ln a} \simeq 
- 4 \frac{\rho_r}{\rho_\phi} - \frac{\Gamma_\phi}{H} \frac{\rho_\phi}{\rho_r} 
\,. 
\ee
The main contribution in the above equation comes from the second
term, which remains constant in this regime with $\Gamma_\phi
\rho_\phi/(H \rho_r) \simeq 2.5$, and then $S_{\phi r} \propto a^{-2.5}$. 
Therefore, when the system
of matter and radiation becomes effectively coupled, the 
entropy perturbation actually decreases due to the relative increase
of radiation coming from matter, and this effect is also seen
in Fig. (\ref{plot1}).   
Obviously, in this simplest model of reheating once the inflaton
completely decays and we are left only with radiation, the relative entropy
perturbation vanishes in any case. We have just integrated the
equations only until the time $\Gamma_\phi \simeq H$, without
continuing the evolution afterwards when all the energy density in
matter becomes radiation. 
However, it is clear that the value of the curvature perturbation,
which becomes practically constant near the beginning of reheating,
will not subsequently be affected by the details 
near the end of reheating and subsequent evolution.

\section{Discussion} 

It is worth pointing out a few subtleties regarding
details of the transition from inflation to
reheating. A detailed description of such transition may be
relevant for a more accurate estimation of the final value of ${\cal
R}$. Here we raise some of the
concerns related to this point, and point out to what extent
our conclusions can be affected. 
During inflation the background evolution equations for the
fields are approximately given by the slow-roll equations,
Eqs. (\ref{dotphi})-(\ref{dotN}); whereas reheating starts with the
fields already oscillating and their energy densities averaged like matter
until they decay into radiation. During the transition from one to
another, the kinetic energies of the fields start increasing and they
cannot be neglected any longer. Indeed it is going to be the $N$ field
which first starts  moving away from the false vacuum once $\phi$
reaches the critical value, with its field-dependent squared mass
becoming negative, and its vev and velocity growing accordingly. Its quantum
fluctuations $Q_N$ will also feel the same instability, and increase
until the effective mass become positive again and the oscillations
begin. Given Eq. (\ref{RQ}), this may mean that during the transition
period it is the $N$ field, 
with both its kinetic energy and the
amplitude of its fluctuations increasing relatively to the others, 
which will end dominating the curvature
perturbation. Still, our conclusion that the Higgs is responsible for
the final value of the adiabatic perturbations holds, although
indirectly. It is through the coupling of the Higgses to the $N$ field
that the quantum fluctuations of the latter are not suppressed during
inflation, as it should correspond to a massive field (see Appendix
A). We also note  that $Q_N/\dot N \simeq Q_h/\dot h$, so that even if
the $N$ kinetic energy dominates before the oscillations begin, we
will end in any case with ${\cal R} \sim H Q_N/\dot N \sim H Q_h/\dot
h$, and our order of magnitude estimation holds. However, this assumes
that the ratio $Q_N/\dot N$ remains constant once the field $N$ start
moving toward the global minimum. This does  not seem unreasonable given that
both the background field and its quantum fluctuations will feel the
same instability in the effective mass. 
However, it may happen that as the fields $N$ and $h$ move
toward the global minimum with increasing speed $\dot N$ and $\dot h$,
their quantum fluctuations still behave as those of a light field
without changing appreciably. This will tend to decrease the final
ratios $Q_N/\dot N$ and $Q_h/\dot h$, and therefore the value of ${\cal
R}$. 

On the other hand, we have started evolving the perturbations
during reheating after the Higgs field decay into radiation. As
mentioned in the previous section, before the Higgs decay the fields
still have time to oscillate several times, such that the energy
density is redistributed among the three oscillating scalar fields. 
During the initial stage of the oscillations, as
far as the three species behave $all$ on average like
matter,  the curvature perturbation ${\cal R}$ remains practically
constant on super-Hubble scales and it is not affected by entropy
perturbations (see Eqs. (\ref{weta}) and ({\ref{dotR}) in Appendix
B). In other words, the change in the curvature perturbation is given by
the large-scale non-adiabatic (entropic) pressure perturbation, which
is negligible as far as the scalar fields behave with similar
effective equation of state. However, this argument does not take into
account the possible parametric amplification of the scalar quantum
fluctuations in a 
background of oscillating fields, a process that in the context of
inflation generically is known as ``preheating''
\cite{preheating1,preheating2}. Through parametric resonance, induced
by the time dependent effective mass term in the evolution equations,
field mode amplitudes can grow exponentially with time within certain
resonance bands in k-space. Thus, preheating become a more effective
mechanism of transferring energy from the oscillating background
fields to quantum fluctuations than the standard perturbative decay of
the fields. The question then is whether it also provides a different
and efficient source for non-adiabaticity, with 
super-Hubble ($k \ll aH$)
fluctuations exponentially amplified during preheating
\cite{kodama2,basset}. This would 
translate into an exponential increase in the curvature perturbation
which would be somehow difficult to keep at the level of the COBE
observational value. As usual, the answer is model dependent.  For
example,  it has been pointed out that for a massive (mass larger than
the Hubble parameter) field during inflation the mechanism is
inefficient, 
because previous to preheating fluctuations are first exponentially
damped during inflation  
\cite{damping}. However, this restriction does not apply to  light
fields during inflation \cite{counterexample}, as in the model 
presented in this paper. Moreover, in hybrid inflation the effect
can be stronger due to the presence of negative effective squared
masses at the end of inflation \cite{us,tachyonic}, which are per se a
source of instabilities in the evolution equations for the fluctuations.     
To which extend the combined  effect on super-Hubble
scales is relevant is then a question of $when$ the resonance ends
\cite{brandenberger}. Sooner or later, this occurs due to the
backreaction effect of the small-scale (sub-Hubble) modes which also
grow resonantly during 
preheating;  in some models it may happen before the curvature
perturbation exceeds the COBE value, depending on the value of the
inflaton coupling constant. Again, the issue is highly model
dependent. Preheating in the present model has been studied previously
in Ref. \cite{us}, but without
neither taking into account the Higgs fluctuations nor metric
perturbations. It was 
shown that the combination of parametric resonance plus tachyonic
instability gives rise to the exponential grow of the inflaton and $N$
field fluctuations, 
and the resonance was stronger on the small-scale perturbations around
$k \sim 0.3 \bar{m}_\phi$. At
the same time and because of that, backreaction effects become
important very quickly after only three oscillations of the fields. 
Given that the tachyonic instability will be still present mainly in
the effective $N$ mass, 
we do not expect this conclusion to change once the Higgs fluctuations
are also taken into account. Even they may speed up the end of the
resonance, keeping under control the growth of the curvature (and
entropy) perturbations. Moreover, an increase due to parametric resonance
might compensate for a potential decrease during the end of the
slow-roll regime. Nevertheless, the final answer
would require the numerical integration of the evolution equations,
for both the background fields and quantum fluctuations, from
inflation to reheating, which is  beyond the scope of the present
publication.  

\section{Spectral index, gaussianity and gravity waves}

In this section we discuss the main predictions
of the mechanism discussed here as compared to both the standard
curvaton mechanism, based on the late-decaying scalar, or
on standard hybrid inflation.

In the original hybrid inflation model based on the NMSSM \cite{steve},
where the inflaton was responsible for curvature perturbations,
we predicted a flat spectrum of curvature perturbations  
with a spectral index indistinguishable from $n=1$.
In the present NMSSM hybrid inflation model, in which the Higgs has large
isocurvature perturbations which get transferred to curvature perturbations
during reheating, we would expect a spectral index controlled by the
Higgs fields instead of the inflaton and given by\footnote{This
prediction for the spectral index is valid when there is no relic
isocurvature perturbation which can significantly affect the CMB power
spectrum.}   
\be
n -1 = 2 \eta_h - 6 \epsilon_\phi \simeq 2 \eta_h \simeq 2m^2_h/(3 H^2).
\label{prediction}
\ee
The prediction for the spectral index in Eq. (\ref{prediction})
involves the soft mass parameter $m_h$ which depends on the details
of the supersymmetry breaking mechanism.
For example in the extra-dimensional model introduced in the
companion paper \cite{vicente}, Higgs soft masses are 
not expected to be much suppressed 
with respect to the Hubble rate of expansion \cite{vicente}.
Using $n=0.93\pm0.13$ \cite{cmb} we have 
an upper bound on the spectral index $n<1.06$ which
leads to a constraint on the soft Higgs mass of $\eta_h < 0.03$. The
inflaton mass on the other hand is only constrained  by the slow-roll
condition $\eta_\phi < 1$, since the inflaton does not contribute 
significantly to curvature perturbations. 
We would therefore expect typical deviations such as 
$\mid n-1 \mid \sim 0.1$. 
If $n=1$ is measured 
very accurately then this would be evidence for the original
inflaton generated curvature perturbations.

In the original single field inflaton model \cite{steve} we predicted
a Gaussian CMB spectrum. 
In the present case we would expect small non-Gaussian
effects, as discussed in Ref. \cite{lyth2}, 
which contribute quadratically to the Higgs energy density, 
\be
\frac{\delta \rho_h}{\rho_h} \simeq 2 \frac{\delta h}{h} +
\frac{(\delta h)^2 }{h^2} \,. \label{deltarhoh}
\ee
Given that the amplitude of the field perturbations at horizon
crossing are of the order of the Hubble parameter, 
$\delta h\sim H_* \sim 10$ MeV, 
but the typical value for the vev of the Higgs is $h \sim 1$ TeV, 
the linear term clearly dominates in
Eq. (\ref{deltarhoh}) given a Gaussian spectrum. The
small non-Gaussian effects can be parametrised by \cite{fnl}
\be
| f_{NL} | \simeq \frac{5}{12} \left| \frac{\delta
\rho_h/\rho_h}{{\cal R}} \right| \sim
\frac{75}{8} |\eta_h | \,,
\ee
where we have used the approximations in
Eqs. (\ref{sphihi}), (\ref{Rf}) and (\ref{deltarhoh}).
Within the approximations involved in the above equations,
and using $\eta_h <0.03$ we then get $\mid f_{NL} \mid < 0.3$, which 
is below the expected upper bound by the PLANCK satellite $\mid
f_{NL}\mid < 5$ \cite{fnl}. Therefore we do not expect significant
(observable) non-Gaussian effects in this model. 

Concerning gravity waves, as is 
typical of hybrid inflation models with a low inflation
scale, the previous model \cite{steve} predicted negligible effects in the CMB
spectrum from gravity waves. In the present approach
we similarly expect small effects of gravity waves in
the CMB spectrum. Gravitational waves are generated with a spectrum
\cite{inflation} 
\be
P_T^{1/2} \simeq 8 \sqrt{2} \frac{H_*}{m_P} \,,
\ee
and the tensor-scalar ratio, from Eqs (\ref{Rf}) and above,  is then
given by 
\be
\frac{P_T}{P_{{\cal R}}} \simeq (120 \pi)^2 \epsilon_h \propto
\left(\frac{h_*}{m_P}\right)^2,
\ee
where the value of the Higgs field at the time of horizon exit is 
typically $h_*\sim 1 TeV$, which is negligible compared to the
reduced Planck mass $m_P$. Therefore we do not expect any observable
effects of gravity waves in the CMB spectrum.

It has been pointed out that in general for the curvaton models
based on a late-decaying scalar
the entropy perturbation originated during inflation might survive the
reheating era as a relic isocurvature perturbation between radiation
and some other present component, say cold dark matter (neutralinos)
or neutrinos \cite{moroi2,lyth2}.
In the framework described in this paper the Higgs fields
can decay into neutralinos and in addition give rise to leptogenesis
during the reheating process \cite{Bastero-Gil:2000jd}.
Therefore the Higgs isocurvature perturbations might be expected
to give rise to relic isocurvature perturbations both in neutralino
dark matter and in baryons, as compared to photons. 
However the Higgs excitations 
decay near the beginning of the reheating era, and any
neutralinos produced at that time are relativistic and share the curvature
perturbation with that of the photons. Therefore once the neutralinos
decouple from the radiation fluid, their curvature perturbation will remain
constant and equal to that of the radiation at the time. On the other
hand throughout reheating the radiation will be coupled to the matter
energy density through the decay of the inflaton field. 
For such a coupled system, 
individual curvature perturbations are not conserved, 
unlike the late-decaying scalar scenario in \cite{lyth1,moroi,others,moroi2,lyth2,lyth3},
and so in our case we expect the isocurvature perturbation 
of the photons to change in a complicated way during the reheating process. 
Thus in our case the final neutralino isocurvature perturbation will
depend on the details of the reheating process, in particular
when the neutralino decouples and becomes non relativistic
during the reheating process. It would be interesting to explore
this quantitatively in a future work.

\section{Summary}

To summarise, we have proposed and discussed a new implementation of 
the curvaton scenario \cite{lyth1,moroi,others,moroi2,lyth2,lyth3} in which 
quantum fluctuations of light scalar fields other than the inflaton can
be responsible for the generation of the primordial anisotropies in
the Universe. Unlike the previous scenarios, we do not assume
a late-decaying scalar which is decoupled from the inflaton field.
Instead we have proposed 
a new mechanism based on supersymmetric hybrid inflation
in which the isocurvature perturbations of a hybrid field
coupled to the inflaton could be transferred to curvature
perturbations during the initial stages of reheating.
We have further suggested that good candidates for such fields
are the Higgs doublets of the supersymmetric standard model,
which may have a flat direction during inflation provided
that the $\mu$ term is generated after inflation.
We have discussed in detail a specific model which implements
this mechanism, based on a 
supersymmetric hybrid inflation model which is a variant of the NMSSM,
based on the superpotential in Eq.(\ref{superpot}).
We have shown that for this particular model our mechanism leads to 
\be
P^{1/2}_{\cal R} \sim \frac{H_*}{2 \pi h_*} \,,
\ee   
which for a typical value of the Higgs vev $2 \pi h_* \simeq 1\, TeV$
gives the correct order of magnitude as measured by COBE
$P^{1/2}_{\cal R} \sim 10^{-5}$. 
This is an important success of both the particular model and
of the proposed mechanism in general. The main prediction is that
the spectral index is expected to deviate significantly from unity
$\mid n-1 \mid \sim 0.1$. If $n=1$ is measured 
very accurately then this would be evidence for the original
inflaton generated curvature perturbations. No observable signals of
non-Gaussianity, or gravity waves are expected in the CMB spectrum.
However there may be observable 
relic isocurvature perturbations between radiation
and some other present component, say cold dark matter (neutralinos)
or neutrinos which depend on the details of reheating and differ
significantly from the expectations of a late-decaying scalar
\cite{moroi2,lyth2}.

To conclude, we have proposed and discussed 
the attractive possibility of having the Higgs field
of the supersymmetric standard model as being
responsible for the large-scale structure
of the Universe, providing a further strong 
link between cosmology and particle physics of the kind
recently emphasised in \cite{Kane:2001rb}.
Eventually, if such a theory as that presented here is realised in nature,
it should be possible, from both laboratory and cosmological measurements, 
to demonstrate the links between particle physics and cosmology 
inherent in such a model \cite{Kane:2001rb}.

\section*{Acknowledgements}

V.D.C and S.F.K. would like to thank PPARC for a Research
Associateship and a Senior Research Fellowship.


\section*{Appendix A} 

The equation of motion for the gauge invariant quantum fluctuation
$Q_\alpha$, with comoving wavenumber $k$, can be written as \cite{taruya}  
\be
\ddot Q_\alpha + 3 H \dot Q_\alpha + \frac{k^2}{a^2}Q_\alpha + \sum_\beta
\left[ V_{\alpha \beta} 
-\frac{1}{a^3 m^2_P} \left( \frac{a^3}{H} \dot \phi_\alpha \dot
\phi_\beta \right)^{\cdot } \right] Q_\beta=0 \,,
\ee
where $V_{\alpha_\beta}= \partial^2 V/\partial \phi_\alpha
\partial \phi_\beta$. For the inflaton and Higgs fields during inflation,
we have $V_{\phi\phi}$, $V_{hh}$, $V_{\phi h} \ll H^2$, and therefore the
fluctuations $Q_\phi$ and  $Q_h$ will be frozen to a constant value
once outside the horizon, $k < aH$, given approximately by the value
at horizon crossing, 
\be
Q_{\alpha *}= \frac{H_*}{\sqrt{2 k^3}} \label{qi}\,,
\ee
for $\alpha=\phi,\,h$. 
On the other hand, neglecting for simplicity the sub-dominant terms coming
from metric contributions, the evolution equation for $Q_N$ reads   
\be
\ddot Q_N + 3 H \dot Q_N + (\frac{k^2}{a^2} + V_{NN} ) Q_N  
 + V_{N\phi} Q_\phi + V_{Nh} Q_h \simeq 0 \,.
\ee 
Like in the case of  the evolution of the background field $N$, 
now the large mass term  $V_{NN} \sim O( \kappa^2 \phi_c^2) \gg H^2$ gives 
rise to oscillations with an amplitude decaying as $a^{-1}$, but
displaced from zero due to the $Q_\phi$ and $Q_h$ terms. 
Therefore, after horizon exit, and 
averaging over the fast oscillations in a Hubble time, the $N$ field
fluctuation will also tend to a constant value given by
\be
Q_N \simeq \frac{V_{N\phi}}{V_{NN}} Q_\phi + \frac{V_{Nh}}{V_{NN}} Q_h \,.
\ee
Using $V_{NN} \simeq 2 \kappa^2 \phi_c^2$, and
$V_{N\phi} \simeq 4 \kappa^2 \phi_c N - \lambda \kappa h^2$, 
$V_{Nh} \simeq 2 \lambda \kappa \phi_c h$, 
we have
\bea
Q_N &\simeq& (2 \frac{N}{\phi_c}-\frac{\lambda h^2}{2 \kappa
\phi_c^2}) Q_\phi + \frac{\lambda h}{\kappa \phi_c} Q_h \nonumber \\
&\simeq& \frac{\lambda h}{\kappa \phi_c} Q_h \,. 
\eea
Therefore, the fluctuation $Q_N$ is suppressed with respect to $Q_h$
by the same 
factor than the background field $N$ is suppressed with respect to
$h$, Eq. (\ref{Nh}), such that,
\be
\dot N \simeq - \frac{\lambda h}{\kappa \phi_c} \dot h \,,
\ee
and then
\be
\frac{Q_N}{\dot N} \simeq - \frac{Q_h}{\dot h} \,.
\ee

\section* {Appendix B}

In this Appendix we summarise our conventions and notation for the
perturbations, and give  the evolution equations for curvature
and entropy perturbations in a multi-component fluid. These can be found in
Refs. \cite{kodama,silvia}. Using linear perturbation theory, 
the equations are given in terms of gauge-invariant quantities
$\Delta_c=\delta \rho_c/\rho$, $V$ and $\eta$ describing the amplitude
perturbation of the 
total density in the comoving frame, velocity and entropy
respectively. In addition we have the relative gauge-invariant
variables between any two components $\alpha$ 
and $\beta$: entropy $S_{\alpha \beta}$ and relative velocity
$V_{\alpha \beta}$.

Gauge-invariant quantities are defined from the original perturbations
in the stress-energy tensor and the metric. 
Linear scalar perturbations of the metric are given by the line element:
\be
ds^2= -(1+2 A) dt^2 + 2 a  B_{i} dx^i dt  \nonumber \\
      + a^2 [ (1-2 \psi) \delta_{ij} + 2 E_{ij}] dx^i
dx^j \,, 
\label{ds2}
\ee
where $\psi$ is the gauge-dependent metric perturbation. Combining
$\psi$, $B$ and $E$, one can define the gauge-invariant variable 
\be
\Phi = -\psi - H a (B- a \dot E) \,,
\ee
which is the curvature perturbation in the longitudinal (or zero-shear)
gauge ($a \dot E_l - B_l=0$). 

Each fluid component is described by a perfect-fluid stress-energy
tensor, with background energy $\rho_\alpha$ and pressure $P_\alpha$. 
Although the total stress-energy tensor is conserved, the
stress-energy tensor of each component may not be conserved 
individually. Therefore, the unperturbed continuity equation for a
given component $\alpha$ is given in general by:  
\be
\dot \rho_\alpha + 3 H (\rho_\alpha + P_\alpha)= C_\alpha\,,
\ee
where a dot denotes derivate with respect to time, and $C_\alpha$ are
source terms  subject to  the total energy-momentum 
conservation constraint $\sum_\alpha C_\alpha=0$. For example, in
Eqs. (\ref{rhophidot}) and (\ref{rhordot}) we have $C_\phi=-C_r=-\Gamma_\phi
\rho_\phi$. For later use, we
define the equation of state for each component,
$w_\alpha=P_\alpha/\rho_\alpha$, and the sound velocity $c^2_{s\alpha}=
\dot P_\alpha/\dot \rho_\alpha$. To simplify notation, we also define
$h_{\alpha}= \rho_\alpha + P_\alpha$.

Perturbations in the stress-energy tensor are given by  $\delta
\rho_\alpha$, $\delta P_\alpha$  
and the momentum perturbation $\delta q_\alpha$ (neglecting the
anisotropic stress). 
Gauge-invariant variables $\delta \rho_{c
\alpha}$, $V_\alpha$ and $\eta_\alpha$ are defined as:
\bea
\delta \rho_{c \alpha} &=& \delta \rho_\alpha + \dot
\rho_\alpha\frac{\delta q}{\rho+P} \,,\\   
V_\alpha &=& - \frac{k}{a} \left(\frac{\delta q_\alpha}{\rho_\alpha +
P _\alpha} - a B + a^2 \dot E \right) \,, \\
P_\alpha \eta_\alpha &=&\delta P_\alpha - c^2_{s \alpha} \delta
\rho_\alpha \,, 
\eea
such that 
\bea
\delta \rho_c &=& \sum_\alpha \delta \rho_{c\alpha}\,, \\
 (\rho + P) V &=& \sum_\alpha (\rho_\alpha+P_\alpha) V_\alpha \,, \\
S_{\alpha \beta} &=& \frac{\delta \rho_{c\alpha}}{\rho_\alpha +
P_\alpha} - \frac{\delta \rho_{c\beta}}{\rho_\beta+P_\beta}
\,,\label{entropy} \\
V_{\alpha \beta}&=& V_\alpha - V_\beta \,.
\eea
The total entropy perturbation (or non-adiabatic pressure
perturbation) $P \eta = \delta P - c^2_s \delta \rho$
can be written  in terms of the individual
entropy perturbations $\eta_\alpha$ for each component and $S_{\alpha
\beta}$ as:
\be
\frac{w}{1+w} \eta = \sum_{\alpha} \frac{h_\alpha}{h} \frac{w_\alpha
\eta_\alpha}{1+w_\alpha} + \frac{1}{2} \sum_\alpha \sum_\beta
\frac{h_\alpha h_\beta}{h^2}(c^2_{s\alpha} -c^2_{s\beta}) S_{\alpha \beta}
+ \frac{\Delta_c}{1+w} \sum_\alpha c^2_{s\alpha}
\frac{C_\alpha}{3 H h} \label{weta}\,.
\ee
Individual entropy perturbations $\eta_\alpha$ vanish for adiabatic
perturbations, but cannot be neglected for example in the case of a
scalar field, in which case they are given by
\be
P_\alpha \eta_\alpha= (1- c^2_{s\alpha}) (\delta \rho_{c \alpha}
-\frac{a}{k} \dot \rho_\alpha (V_\alpha-V) ) \,.
\ee

The comoving curvature perturbation is  defined by:
\be
{\cal R} =  \psi - \frac{H}{\rho+P} \delta q 
           = -\Phi + \frac{a H}{k} V 
= -\frac{3}{2} \left(\frac{a H}{k}\right)^2 
\Delta_c + \frac{a H}{k} V  \label{curvature}\,,
\ee
where in the last equality  we have used the relation between $\Phi$ and
the comoving total 
density perturbation through the energy constraint given by
\be
\Phi=\frac{3}{2} \left( \frac{a H}{k} \right)^2 \Delta_c \,.
\ee
On the other hand, the curvature perturbation on constant-density
hypersurfaces is given by
\be
\zeta= -\psi - H \frac{\delta \rho}{\dot \rho} \,,
\ee
which is related to the comoving curvature perturbation ${\cal R}$ by
\be
-\zeta=  {\cal R} - \frac{2 \rho}{9(\rho + P)}
\left(\frac{k}{aH}\right)^2 \Phi \,.
\ee
Therefore, for super-Hubble scales, $k \ll aH$,
$\zeta$ and ${\cal R}$ are approximately the same.

We note that we have followed Refs. \cite{silvia,kodama} in defining
the entropy perturbation $S_{\alpha \beta}$. An alternative
definition related to the individual curvature perturbations
$\zeta_\alpha$, used in Refs. \cite{lyth1,lyth2}, is given by
\be
\tilde{S}_{\alpha \beta} = - 3 H \left( \frac{\delta \rho_\alpha}{\dot
\rho_\alpha} - \frac{\delta \rho_\beta}{\dot \rho_\beta} \right)= 3
(\zeta_\alpha - \zeta_\beta )  \,.
\label{szeta} 
\ee
In terms of $\tilde{S}_{\alpha \beta}$, the total entropy perturbation
is then given by:
\be
\frac{w}{1+w} \eta = \sum_{\alpha} \frac{h_\alpha}{h} \frac{w_\alpha
\eta_\alpha}{1+w_\alpha} + \frac{1}{2} \sum_\alpha \sum_\beta
\frac{h_\alpha h_\beta}{h^2}(c^2_{s\alpha} -c^2_{s\beta}) \tilde{S}_{\alpha
\beta}\,. 
\ee
Nevertheless, for a system of uncoupled fluids, Eqs. (\ref{entropy}) and
(\ref{szeta}) become identical. $S_{\alpha \beta}$ and
$\tilde{S}_{\alpha \beta}$ differ by terms proportional to the source terms
$C_\alpha$ when there is energy transfer between
components. For example, for a 2-component system, their relation can
be written as
\be
(1- q_\alpha) (1-q_\beta) \tilde{S}_{\alpha \beta} = S_{\alpha \beta}
+ (q_\alpha- q_\beta) \frac{\delta \rho_c}{\rho+P} \,, 
\ee
where $ 1-q_\alpha=-\dot \rho_\alpha/(3 H h_\alpha)$.

The evolution equations for the total density and velocity
fluctuations in a flat Universe are given by:
\bea
\Delta_c^{\prime} &=& 3 w \Delta_c - (1+w) \frac{k}{aH} V \,,
\label{rhoc}
\\
V^{\prime}&=& - V - \frac{3}{2} \left(\frac{a H}{k} \right) \Delta_c
+ \frac{k}{aH} \left( \frac{c^2_s}{1+w} \Delta_c +
\frac{w}{1+w} \eta \right) \,,
\label{Vprime}
\eea
where prime means derivative with respect to $\ln a$. 
Thus, taking the derivative in Eq. (\ref{curvature}) and using
Eqs. (\ref{rhoc}-\ref{Vprime}),  the evolution  
equation for the comoving curvature perturbation reads 
\bea
{\cal R}^{\prime} &=& (1 + 3 w) \frac{3}{2} \left(\frac{a H}{k}\right)^2 
\Delta_c -\frac{3}{2} \left(\frac{a H}{k}\right)^2 
\Delta_c^\prime 
-\frac{1}{2} (1+ 3 w) \frac{a H}{k} V  + \frac{a H}{k} V^\prime \\  
&=& \frac{c^2_s}{1+w} \Delta_c + \frac{w}{1+w} \eta 
                  = \frac{2}{3}
\left(\frac{k}{aH}\right)^2 \frac{c^2_s}{1+w} \Phi + \frac{w}{1+w} \eta \,.
\label{dotR}
\eea

Finally, for a two-component fluid, such that $C_\alpha=-C_\beta$, the
equations for the relative fluctuations $S_{\alpha \beta}$ and 
$V_{\alpha \beta}$ are given by:
\bea
S_{\alpha \beta}^{\prime}&=& -\frac{k}{a H} V_{\alpha \beta} 
- 3 \left( \frac{w_\alpha \eta_\alpha}{1+w_\alpha} -\frac{w_\beta
\eta_\beta}{1+w_\beta} \right) \nonumber \\
& &- \frac{C_\alpha}{H h}
\left(\frac{h_\beta}{h_\alpha}(1+c^2_{s\alpha} )-\frac{h_\alpha}{h_\beta}(1
+c^2_{s\beta}) \right)  S_{\alpha \beta}   
- \frac{C_\alpha}{H}\frac{h}{h_\alpha h_\beta} \frac{ w}{1+w}\eta
\nonumber \\
& &-\frac{C_\alpha}{H} \frac{h}{h_\beta h_\alpha}(1+ c^2_s +
\frac{h_\alpha}{h} c^2_{s\beta} +
\frac{h_\beta}{h} c^2_{s\alpha}) \frac{\Delta_c}{1+w}
+\frac{C_\alpha}{H} \frac{h}{h_\alpha h_\beta} E_{c\alpha} \,,
\label{salpha} \\  
V_{\alpha \beta}^{\prime} &=& (3 c^2_{s \alpha} \frac{h_\beta}{h} +
3 c^2_{s\beta} \frac{h_\alpha}{h} - 1) V_{\alpha \beta} 
+ \frac{k}{aH} (c^2_{s\alpha} \frac{h_\beta}{h} + c^2_{s\beta}
\frac{h_\alpha}{h} ) S_{\alpha \beta} \nonumber \\
& & + \frac{k}{aH} \left( \frac{w_\alpha \eta_\alpha }{1+w_\alpha} 
-\frac{w_\beta \eta_\beta}{1+w_\beta}  \right) 
+ \frac{k}{aH} (c^2_{s\alpha}- c^2_{s\beta}) \frac{\Delta_c}{1+w} \nonumber
\\
& & - \frac{C_\alpha}{H h} \left( (1+c^2_{s\alpha} )\frac{h_\beta}{h_\alpha} -
(1+c^2_{s\beta} )\frac{h_\alpha}{h_\beta} \right) V_{\alpha \beta}
\label{valpha} \,, 
\eea 
where $E_{c\alpha}=E_{c\beta}$ are the gauge invariant perturbations of
the energy transfer term in the comoving frame. For the simplest case
$C_\alpha= \Gamma_\alpha \rho_\alpha$ then $E_{c\alpha}=\delta
\rho_{c\alpha}/\rho_\alpha$.  

\section*{Appendix C}

For the scenario considered in this paper, we can take the component
``$\alpha$'' to be the oscillating fields behaving like matter
($w_\alpha=c_{s\alpha}$=0),``$\beta$'' then refers to the radiation
($w_\beta=c_{s\beta}$=1/3) initially coming from the Higgs decay
products, and $C_\alpha= C_\phi=-\Gamma_\phi \rho_\phi$. Then, the
evolution Eqs. (\ref{dotR}), (\ref{salpha}) and (\ref{valpha}) for the
curvature ${\cal R}$, $S_{\phi r}$ and 
$\hat{V}_{\phi r}=(H a/k) V_{\phi r}$ reduce to:
\bea
{\cal R}^{\prime} &=& \frac{\Delta_c}{1+w} \left( \frac{\rho_\phi+ 4 \rho_r}{3
\rho_\phi + 4 \rho_r} \right)+ \frac{8 \rho_\phi \rho_r}{(3 \rho_\phi + 3
\rho_r)^2} S_{\phi r} + \frac{36 \rho_\phi \rho_r}{(3 \rho_\phi + 3
\rho_r)^2}\left( 1 + \frac{\Gamma_\phi}{3H} \right)
\hat V_{\phi r} \,, \label{rphir}\\
S_{\phi r}^\prime &=& - \left(\frac{k}{a H}\right)^2 \hat V_{\phi r} - 3
\frac{\Delta_c}{1+w} - \frac{12 \rho_r}{ 3 \rho_\phi + 4 \rho_r}
S_{\phi r} - \frac{36 \rho_r}{3 \rho_\phi + 4 \rho_r} \hat V_{\phi r} \nonumber \\
& & - \frac{\Gamma_\phi}{H} \frac{\rho_\phi}{\rho_r} \left( \frac{3
\rho_\phi + \rho_r}{ 3 \rho_\phi + 4 \rho_r} \right) S_{\phi r} 
+ \frac{\Gamma_\phi}{H} \frac{9 \rho_\phi}{ 3 \rho_\phi+ 4 \rho_r}
\left( 1 - 4 \frac{\rho_r}{\rho_\phi} + \frac{\Gamma_\phi}{3 H}
\right) \hat V_{\phi r} \nonumber \\
& & + \frac{\Gamma_\phi}{H} \frac{ \rho_r + 3 \rho_\phi}{3 \rho_r}
\frac{\Delta_c}{1 + w} \,, \label{sphir}\\ 
{\hat V}_{\phi r}^\prime &=& -\frac{1 + 3 w}{2} \hat V_{\phi r}+
\frac{8 \rho_r}{ 3 \rho_\phi + 4 \rho_r} \hat V_{\phi  
r} + \frac{2}{3} \frac{\Delta_c}{1+w} +
\frac{ 4 \rho_r + \rho_\phi}{3 \rho_\phi + 4 \rho_r} S_{\phi r} \nonumber \\
& & + \frac{\Gamma_\phi}{H} \left( \frac{8 \rho_r^2 - 3
\rho_\phi^2}{\rho_r ( 3 \rho_\phi+ 4 \rho_r)}\right) \hat V_{\phi r}
\,. \label{vphir} 
\eea
For super-Hubble scales 
such that $k \ll aH$, we can further simplify the system by neglecting
the contribution of the terms $\propto (k/aH)^2$, and that of
$\Delta_c= 2 (k/a H)^2 \Phi/3$. 
In order to understand the qualitative behaviour of the solutions to
the above system of equations, we can focus on the evolution
of the curvature ${\cal R}$ and $S_{\phi r}$, without taking into
account the terms due to $\hat V_{\phi r}$. From the evolution
equations we can see 
that $S_{\phi r}$ and $\hat V_{\phi r}$ will behave approximately the
same, so both of them will affect the curvature ${\cal R}$  in a
similar way; then the main effect is obtained by keeping
only those term due to $S_{\phi r}$ in the evolution equations. We
have checked that the numerical solution  of
Eqs. (\ref{rphir}-\ref{vphir}) is well approximated by that of:
\bea
{\cal R}^{\prime} & \simeq & \frac{8 \rho_\phi \rho_r}{(3 \rho_\phi + 3
\rho_r)^2} S_{\phi r} \label{rphir2}\\
S_{\phi r}^\prime & \simeq & - \frac{12 \rho_r}{ 3 \rho_\phi + 4 \rho_r}
S_{\phi r} 
- \frac{\Gamma_\phi}{H} \frac{\rho_\phi}{\rho_r} \left( \frac{3
\rho_\phi + \rho_r}{ 3 \rho_\phi + 4 \rho_r} \right) S_{\phi r} 
\,. \label{sphir2} 
\eea
In order to solve analytically the equations, we can distinguish two regimes
depending on the evolution of the background energy densities
$\rho_\phi$ and $\rho_r$, as it can be seen in Fig. ({\ref{plot1}): (a) from
the beginning of the reheating period at $a_0$, with $\rho_r(a_0) \simeq
\rho_\phi(a_0)$, up to say $a_1$, the decay
products coming from the singlets have no effect on the radiation, 
and both components, matter and radiation,  behave as if they were
decoupled, with $\rho_r\propto a^{-4}$ and $\rho_\phi \propto
a^{-3}$; (b) from $a_1$ to the end of reheating,  
the singlets decay products start contributing to the background
radiation, with $\rho_r \propto a^{-3/2}$ and $\rho_\phi \propto
a^{-3}$.  Therefore, for  $a_0 \leq a \leq a_1$, Eqs. (\ref{rphir2}) and
(\ref{sphir2}) can be solved neglecting 
the terms proportional to $\Gamma_\phi/H$, and we
obtain for ${\cal R}$ and $S_{\phi r}$: 
\bea
{\cal R} &\simeq &{\cal R}(a_0)+ \frac{8}{343} S_{\phi r}(a_0) \left( 5
-\frac{2 a_0^2}{a^2} - \frac{ 3 a_0}{a}\right) \,, \\
S_{\phi r} &\simeq& \left( \frac{4 + 3 a/a_0}{7 a/a_0} \right)^3
S_{\phi r} (a_0) \,,
\eea
where ${\cal R}(a_0) \ll S_{\phi r}(a_0)$ are the initial values of the
perturbations taken as those at the end of inflation. It can be seen that very
quickly ${\cal R}$ will tend 
to a constant value with ${\cal R} \simeq S_{\phi
r}(a_0)/10$. Similarly, the entropy perturbations go to $S_{\phi
r} \simeq 0.08 \,S_{\phi r}(0)$. That is, entropy perturbations act as a
source for the curvature perturbations only at the beginning, as far
as the ratio $\rho_r/\rho_\phi$ is not suppressed. Afterwards,
$\rho_r \ll \rho_\phi$, and ${\cal R}$ remains constant.   

Entropy perturbations will not further affect the evolution of the
curvature even when the background equations become coupled at $a_1$,
and the effects of $\Gamma_\phi$ cannot be neglected. At this point
we can still assume that the energy density is dominated by that of the
singlets, behaving like matter, with $H \propto a^{-3/2}$. Therefore,
from $a_1$ onwards, the combination $\Gamma_\phi \rho_\phi/(H \rho_r)$
remains constant until the singlets completely decay. Numerically we
find $\Gamma_\phi \rho_\phi/(H \rho_r) \simeq 2.5$, independently of
the value of $\Gamma_\phi$. Taking this
into account, we can now solve Eq. (\ref{sphir2}) including the
$\Gamma_\phi$ term:
\be
S_{\phi r} ( a> a_1) \simeq 0.07 \,S_{\phi r}(a_1) \left( \frac{a_1}{a}
\right)^{2.5}\,.
\ee
That is, the relative perturbation between matter (singlets) and radiation
decreases. With this result, we can see that the source term in the evolution
equation for the curvature perturbation,
\be
{\cal R}^\prime \simeq \frac{8 \rho_r}{3 \rho_\phi} S_{\phi r}
\propto \left(\frac{a}{a_1}\right)^{-2} \,,
\ee
also decreases in time and it is no longer  effective, even if by the end
of the reheating period $\rho_r \simeq \rho_\phi$ again. 

The evolution of the perturbations plotted in Fig. (\ref{plot1}) is
the result of the numerical integration of the system of
Eqs. (\ref{rphir}-\ref{sphir}), with initial conditions at the end of
inflation \cite{silvia}:
\bea
{\cal R}|_i &\simeq& \frac{H_*}{\dot \phi_*} Q_{\phi_*}  \simeq -\frac
{ Q_{\phi_*}}{ \eta_\phi \phi_*} \,,\\
S_{\phi r}|_i &\simeq& \frac{8}{3} \frac{H_*}{\dot h_*} Q_{h_*} \simeq
-\frac{8}{3} \frac{Q_{h_*}} {\eta_h h_*} \,, \\
\frac{a H}{k} V_{\phi r}|_i&\simeq& -\frac{1}{3} S_{\phi r}|_i \,,
\eea
Given that for the amplitude of the field perturbations at the time of
horizon crossing we have $Q_{\phi *} \simeq Q_{h *}\simeq H_*/\sqrt{2
k^3}$, it is more convenient  to normalise  the above perturbations
with respect to the initial value of the entropy perturbation $S_{\phi
r} |_i$, such that: 
\bea
\left. \frac{{\cal R}}{S_{\phi r}} \right|_i &\simeq& 
-\frac { 3 \eta_h h_*}{8\eta_\phi \phi_*} \,,\\
\frac{a H}{k} \left. \frac{V_{\phi r}}{S_{\phi r}} \right|_i &\simeq&
-\frac{1}{3} \,.
\eea

\end{document}